\documentclass[referee]{aa} 
\usepackage{graphicx}
\def\DV{{\textit{de Vaucouleurs}}}
\def\Se{{\textit{S\'ersic}}}
\begin{document}

\title{Exact solutions for the spatial de Vaucouleurs and S\'ersic 
laws and related quantities}
\titlerunning{Exact solutions...}

\author{
        A. Mazure \inst{1}
\and
        H.V. Capelato\inst{2}
        }

\institute{
 LAM  Traverse du Siphon, BP 8, 13376,  Marseille Cedex 12, France\\
\and
 INPE/DAS  CP 515, S. J. Dos Campos SP 12201-970, Brazil \\
             }

\offprints{A. Mazure}

\date{Received 20 September 2001; Accepted 23 November 2001}

\abstract{
Using the {\it{Mathematica}} package,
we find exact analytical expressions
for the so-called de-projected \DV\ and \Se\ laws as well as for
related spatial (3D) quantities  ~-~ such the mass, gravitational potential, the total
energy and the central velocity dispersion ~-~ generally involved in astronomical
calculations expressed in terms of  the Meijer G functions.
}

\maketitle


\section{Introduction}

Dynamical studies of astronomical systems like Elliptical Galaxies or
Clusters of Galaxies involve the de-projection of observed (projected 
on the sky) quantities like  surface brightness profiles, 
numerical density profiles,  velocity dispersion  profiles etc.The 3D profiles obtained  
are then used to derive e.g the total luminosity (or mass) of the system 
or the 
gravitational potential  and are used in  the Jeans equation 
which is then resolved to get for instance the kinematics of the system.

The \DV\ profile (de Vaucouleurs 1948) and its generalization by the 
\Se\ law  (S\'ersic 1968), is one of the most often used laws 
particularly in the study of Elliptical Galaxies.

Unhappily these laws have so far  lead to 
non-analytical de-projection (i.e spatial) quantities. Efforts have been
made in the last decades to provide either numerical tables (Poveda 
et 
al 1960, Young 1976) or approximations and asymptotic expressions 
(Mellier and  Mathez 1987, Ciotti 1991,  Graham and Colless 1997, 
Ciotti and
Bertin, 1999, Marquez et al. 2001).

Here we give analytical exact expressions for 3D quantities  
usually derived when using the \DV\  or \Se\ laws.


\section{Principles}

The classical \DV\ and \Se\ laws express the dependence on the 
projected central 
distance $R$ of for instance the Luminosity Intensity $I(R)$ of an 
Elliptical Galaxy.

\vspace {3mm}

The \DV\ law relates the Intensity $I(R)$ to the central one, $I(0)$, 
by:
  
\vspace {6mm}

\begin{equation}
I(R)/I_0  = \exp\left[{-7.66925\,[(R/R_e)^{\frac{1}{4}}-1]} \right]
\label{eq:devauc2}
\end{equation}

\vspace {6mm}

and the \Se\ profile, which is a generalization of the \DV\ profile, 
is written:

\vspace {6mm}

\begin{equation}
I(R)/I_0  = \exp\left[{-b(m)\,[(R/R_e)^{\frac{1}{m}}-1]} \right]
\label{eq:sersic2}
\end{equation}

\vspace{6mm}

The parameter $b(m)$ is determined from the definition
of the effective radius $R_e$, which is the projected radius inside 
which
the projected luminosity (or mass)  equals half of the total
luminosity (or mass).  The \DV\ law   is recovered  for $b(4) = 
7.66925$. 
 
\vspace {3mm}

The requested 3D profiles are  related to the derivative of 
the 
projected profiles by the usual Abel Integral written here for the 3D 
density 
profile $n(r)$:

\vspace {3mm}

\begin{equation}
n(r) =   - \frac{1}{\pi}\int 
_{r}^{\infty}\frac{dI}{dR}\frac{1}{\sqrt{R^2-r^2}}\,dR    
\label{eq:deproj}
\end{equation}

\vspace {3mm}

Except for some particular cases 
there is no known exact expression for theses integrals, in
particular in the case of the \Se\ ( \DV\ )  
profiles.

However, using {\sl{Mathematica }} we succeeded in obtaining
exact analytical expressions for such integrals that involve  the
\textit{ Meijer G functions}.  Physical quantities such as the  spatial 
luminosity or mass profiles, the gravitational force, the gravitational 
potential and energy, which are combinations or integrals of the above 
functions, have also analytical expressions involving \textit{Meijer G 
functions}. Some of these expressions are given below and numerical 
evaluations are compared to previous numerical calculations.

We first give the classical definition of the \textit{Meijer G 
functions} together with some of their properties which will be useful 
in understanding the results given by {\sl{Mathematica}}. 

\section{The Meijer functions}

These functions are defined as  integrals of products of $\Gamma$ 
functions.
The {\sl{generalized}} Meijer G function
is defined as:
\begin{eqnarray}  
&&\!\!\!\!\!G^{k,l}_{p,q}\left(z,r|^{\alpha_1 \cdots 
\alpha_l,\alpha_{l+1} \cdots \alpha_p}_{\beta_1 \cdots 
\beta_k,\beta_{k+1} \cdots \beta_q}\right) =  \nonumber \\
&& = \frac{r}{2\pi i}  \int 
{\frac{[\Gamma(1-\alpha_{1}-ru)\ldots\Gamma(1-\alpha_{l}-ru)\Gamma(\beta_{1}+ru) 
\ldots 
\Gamma(\beta_{k}+ru)]}{[\Gamma(\alpha_{l+1}+ru)\ldots\Gamma(\alpha_{p}+ru)\Gamma(1-\beta_{k+1}-ru) 
\ldots\Gamma(1-\beta_{q}-ru)]}\,z^{-u} du}
\label{eq:meijer} 
\end{eqnarray}
(Gradshteyn \& Ryzhik, 1980; \cite{mathbook}; for a collection of formulae
 related to the Meijer G functions cf.  http://functions.wolfram.com/Hypergeometric
Functions/MeijerG/).

\vspace{3mm}

The case $r = 1$ defines the {\sl{standard}} Meijer function,
$G^{k,l}_{p,q}\left(z|^{\{\alpha_1 \cdots \alpha_l\},\{\alpha_{l+1} \cdots 
\alpha_p\}}_{\{\beta_1 \cdots
\beta_k\},\{\beta_{k+1} \cdots \beta_q\}}\}\right) $, which is the form we 
will be dealing
with in the rest of this work.  In the {\sl{Mathematica}} {\verb=StandardForm=} notation
 it  writes as:
\begin{eqnarray}
 {\rm Meijer}G\Big[\{\{\alpha_1 & \cdots &\alpha_l\}\{\alpha_{l+1} \cdots 
\alpha_p\}\}\{\{\beta_1 \cdots \beta_k\}\{\beta_{k+1} \cdots 
\beta_q\}\}, z \Big]  \equiv\nonumber \\
& \equiv & G^{k,l}_{p,q}\left(z \Big{|}^{\big{\{\alpha_1 \cdots \alpha_l\},\{\alpha_{l+1} \cdots 
\alpha_p\}}}_{\big{\{\beta_1 \cdots \beta_k\},\{\beta_{k+1} \cdots \beta_q\}}}\right)  \nonumber 
\end{eqnarray}
For clarity we will keep both these notations, although suppressing the suffix ``Meijer''. 
Notice that  in these formulae, the empty case: $\{~\}$ means that the corresponding coefficients
are not defined and thus do not  exist. This may occur, for instance, when $k = q$, or $l=p$, or
else when one of those indices are null. An example is given by Eq. (\ref{eq:caso1}) below.

The following identity may be easily obtained by a substitution of 
the integration variable $u \rightarrow u + c$, 
where $c$ is a constant, in Eq. (\ref{eq:meijer}):
\begin{equation}
G^{k,l}_{p,q}\left(z \big{|}^{\{\alpha_1 \cdots \alpha_l\},\{\alpha_{l+1} \cdots 
\alpha_p\}}_{\{\beta_1 \cdots \beta_k\},\{\beta_{k+1} \cdots \beta_q\}}\right) \equiv z^{-c} 
G^{k,l}_{p,q}\left(z\big{|}^{\{\alpha_1 + c \cdots \alpha_l + c\},\{ \alpha_{l+1} + c \cdots 
\alpha_p + c\}}_{\{\beta_1 + c \cdots \beta_k +  c \},\{\beta_{k+1} + c \cdots \beta_q + c\}}
\right)
\label{eq:prop1}
\end{equation}

The moments of the Meijer function are expressible in terms of the 
higher-order Meijer functions:

\begin{equation}
\int{z^{\xi} G^{k,l}_{p,q}\left(z \big{|}^{\{\alpha_1 \cdots \alpha_l\},\{\alpha_{l+1} \cdots 
\alpha_p\}}_{\{\beta_1 \cdots \beta_k\},\{\beta_{k+1} \cdots \beta_q\}}\right) dz} =
z^{1 + \xi} G^{k,l + 1}_{p + 1,q + 1}\left(z\big{|}^{\{\alpha_1 \cdots 
\alpha_l, ~~~-\xi~~\},\{\alpha_{l+1} \cdots \alpha_p\}}_{\{\beta_1 \cdots 
\beta_k\},\{ -(1 + \xi),\beta_{k+1} \cdots \beta_q\}}\right)
\label{eq:prop2}
\end{equation}
This may be straightforwardly demonstrated by inverting the order of 
the integrations.

In some particular cases the Meijer G functions may be expressed
in term of more classical special functions.  As an example we give below the case
of the Meijer function $G^{2,0}_{0,2}\left(z|\beta_1, \beta_2\right)$ (c.f.  http://functions.wolfram.
com/07.09.03.0330):

\begin{equation}
G^{2,0}_{0,2}\left(z|\beta_1, \beta_2\right)  \equiv 
 G^{2,0}_{0,2}\left(z\big{|}^{~~~~~~\{\},\{\}}_{\{\beta_1, \beta_2\},\{\}}\right) = 2 z^{(\beta_1 + \beta_2)/2}
K_{\beta_1 - \beta_2}(2 \sqrt{z})
\label{eq:caso1}
\end{equation}

where $K_{\tau}(x)$ is the modified Bessel function of $\tau-$ order.

\section{The 3D laws}

In the following,  we give  the analytical expressions using the
{\sl{Mathematica}} formalism for the $G$ Functions. 

Let first  start by introducing some useful dimensionless quantities.
The dimensionless 2D and 3D  $x$ and $s$ radial distances are 
expressed in terms 
of $R_{e}$ as:

\begin{equation}
x \equiv R/R_e  ~~;~~  s \equiv r/R_e
\label{eq:def_x-s}
\end{equation}

The dimensionless 2D and 3D profiles are also defined as:

\begin{equation}
i(x) \equiv I(R)/I_0  ~~{\rm{and}}~~ \nu(s) \equiv n(r) 
\frac{R_e}{I_0}   
\label{eq:def_i-nu}
\end{equation}

which lets equations (\ref{eq:devauc2}) and (\ref{eq:sersic2}) 
transform in the following reduced form:

\vspace {3mm}

\begin{equation}
i(x) = \exp{\left[-b(m)(x^{\frac{1}{m}} -1)\right]}
\label{eq:profadim}
\end{equation}

\vspace {3mm}
where the \DV\ law is obtained for  $m=4$. 

The de-projection integral  (\ref{eq:deproj}) also transforms to:

\vspace {3mm}

\begin{equation}
\nu(s) =-\frac{1}{\pi} 
\int_{s}^{\infty}{\frac{di}{dx}\frac{1}{\sqrt{x^2 - s^2}}}\,dx
\label{eq:deprojadim}
\end{equation}

\subsection{ De-projection of the de Vaucouleurs law}

We give here some detailed results for the case of the \DV\ law as 
an example and give more general results in the next section using 
the \Se\ ~law. Integrating  the  preceding equations formally using  
{\sl{Mathematica}}, we first derive  the  expression for the 3D 
profile $n(r)$. We  will then calculate the  luminosity (or mass) 
profiles as well as  the gravitational potential and the gravitational
energy. 

Only numerical estimations or asymptotic behaviors were given before
(Poveda et al, 1960; Young, 1976; Mellier \& Mathez, 1987) so we will
compare our results with those provided by Young for the spatial 
density and for the luminosity (or mass).

The  basic ingredient to obtain the $\nu(s)$ profile is the 
derivative of the \DV\ law,  equation (\ref{eq:devauc2}), relative 
to the projected dimensionless distance, $x$, which is written:

\vspace {3mm}

\begin{equation}
\frac{di}{dx}=-\frac{b}{4x^{\frac{3}{4}}}\,\exp{[-b(x^{\frac{1}{4}}-1)]} 
~~{\rm{with}}~~ b=7.66925
\label{eq:devauc2prime}
\end{equation}

\vspace {3mm}

Integrating equation (\ref{eq:deprojadim}),
 the spatial density $\nu(s)$ expressed in terms of the dimensionless 
3D radial distance 
 $s$ is then given by:

\vspace {3mm}

\begin{equation}
\nu(s) = 6.23828  s^{-7/4}\,G(\{ \{ \} ,\{ \} \} ,\{ \{ 
\frac{1}{2},\frac{5}{8},\frac{3}{4},\frac{7}{8},\frac{7}{8},1,\frac{9}{8},\frac{5}{4}\},\{ 
\} \} , 0.713351\,s^2) 
\label{eq:devauc_nu}
\end{equation}

\vspace {3mm}

One then obtains the luminosity (or mass) spatial profile defined by:

\vspace {3mm}

\begin{equation}
M(s) = 4 \pi  \int _{0}^{s}  s'^2 \nu(s') ds'
\label{eq:def_mass}
\end{equation}

\vspace {3mm}

which with a new formal integration gives:

\vspace {3mm}

\begin{equation}
   M(s) = 39.1962\,s^{\frac{5}{4}}\,G(\{ \{ \frac{3}{8}\} ,\{ \} \} 
,\{ \{ \frac{1}{2},\frac{5}{8},\frac{3}{4}, 
\frac{7}{8},\frac{7}{8},1,\frac{9}{8},\frac{5}{4}\} , \{ -\left( 
\frac{5}{8} \right) \} \} ,0.713351\,s^2)    
\label{eq:devauc_M}
\end{equation}

\vspace {3mm}

 The gravitational potential is defined by:

\begin{equation}
\Psi(s) = \int _{0}^{s} \frac{M(s')}{s'^2} ds'
\label{eq:def_grav_pot}
\end{equation}
which  gives:
%
\begin{eqnarray}
&&\!\!\!\!\!\Psi(s) =  19.5981\,s^{\frac{1}{4}} \ast \nonumber  \\
&&\;\;\;G(\{ \{ \frac{3}{8},\frac{7}{8}\} ,\{ \} \},\{ 
\{\frac{1}{2},\frac{5}{8},\frac{3}{4},\frac{7}{8}, 
\frac{7}{8},1,\frac{9}{8},\frac{5}{4}\} ,\{ -\left( \frac{5}{8} 
\right) , -\left( \frac{1}{8} \right) \} \} ,0.713351\,s^2) 
\label{eq:pot_dv}
\end{eqnarray}

\vspace {3mm}

The values tabulated by Young (1976) are recovered defining the new 
potential:
$$
\Psi'(s) = 1- \Psi(s)/\Psi(\infty) 
$$
such that
$$
\Psi'(0) = 1 ~~{\rm{and}}~~ \Psi'(\infty)) = 0 
$$

\vspace {3mm}

In  table 1, we compare the values given with the above  expressions 
(normalized by the factor $exp(b) (\pi 8!/b^{8})$) to 
the numerical values obtained by Young for $s$ ranging from $10^{-6}$ 
to 
$10$.

\vspace {3mm}

\begin{table}[htb]
\caption[]{The spatial density and mass profiles for the de 
Vaucouleurs profile }
\tabcolsep=0.6\tabcolsep
\begin{tabular}{l c c c c}
\hline
      $s$ &        $\nu(s)$ & $\nu_{Young}(s)$      &             
$M(s)$  & $M_{Young}(s)$     \\
\hline
1$\cdot10^{-6}$ & 2.5553$\cdot10^6$    & 2.5535$\cdot10^6$    & 
1.4730$\cdot10^{-11}$ & 1.4717$\cdot10^{-11}$ \\
1$\cdot10^{-5}$ & 3.5797$\cdot10^5$    & 3.5786$\cdot10^5$    & 
2.1130$\cdot10^{-9}$  & 2.1122$\cdot10^{-9}$  \\
1$\cdot10^{-4}$ & 4.2189$\cdot10^4$    & 4.2183$\cdot10^4$    & 
2.5959$\cdot10^{-7}$  & 2.5954$\cdot10^{-7}$  \\
1$\cdot10^{-3}$ & 3.7044$\cdot10^3$    & 3.7042$\cdot10^3$    & 
2.4545$\cdot10^{-5}$  & 2.4544$\cdot10^{-5}$  \\
1$\cdot10^{-2}$ & 1.9679$\cdot10^2$    & 1.9679$\cdot10^2$    & 
1.4961$\cdot10^{-3}$  & 1.4960$\cdot10^{-3}$  \\
1$\cdot10^{-1}$ & 4.4047               & 4.4047               & 
4.4102$\cdot10^{-2}$  & 4.4102$\cdot10^{-2}$  \\
1.             & 2.1943$\cdot10^{-2}$ & 2.1943$\cdot10^{-2}$ & 
4.1536$\cdot10^{-1}$  & 4.1536$\cdot10^{-1}$  \\
10.            & 7.8166$\cdot10^{-6}$ & 7.8165$\cdot10^{-6}$ & 
9.4308$\cdot10^{-1}$  & 9.4308$\cdot10^{-1}$  \\ 
\hline
\label{devaucvalues}
\end{tabular}
\end{table}

\vspace {3mm}

\subsection{The de-projection of the S\'ersic law}

\vspace{3mm}

We give now more general expressions for the  3D profile derived from 
the \Se\ law.
This law is parametrized by an index $m$ and a parameter $b(m)$.
The \Se\ law for  $m=1$ corresponds to a 3-D \textit{Exponential}
profile, already discussed by Fuchs \& Materne (1982) and for  $m=4$ to 
the usual \textit{de  Vaucouleurs} law, treated in Section 2 above. In general, 
$b(m)$ is found as a solution of
equation (\ref{eq:bm}) given in the Appendix. Values of $b(m)$, (as well as  $\log L(m)$)
 have also been given before by Ciotti,  Graham \& Colless  and Ciotti \& Bertin. 
We give in the  Appendix, values of $b(m)$ calculated using
{\sl{Mathematica}} for the  range $m=1$ to $m=15$ and compared to 
those of Ciotti \& Bertin.

\vspace{3mm}

 The spatial density expressed in terms of the dimensionless 3D 
 distance $s$ can be  written  as  (see e.g. Ciotti, 1991; Graham \& 
Colless, 1997):

\vspace{3mm}

\begin{equation}
\nu(s) = \frac{be^b}{\pi} s^{(1/m-1)}
\int _{0}^{1.}\exp{(-\frac{b\,s^{\frac{1}{m}}}{t}})\,\frac{1}{t^2\,
\sqrt{t^{-2\,m}-1}}\,dt
\label{eq:deprojsersic}
\end{equation}
where $b$ means $b(m)$

\vspace{3mm}

  Again this integral is found in terms of { Meijer G Functions 
} which 
  reduces to a\textit{ Bessel } function  ($ K_{0}$) for $m=1$. We 
give below the various 
  expressions for various values of $m$ and $b(m)$. Defining 
auxiliary constants:

\begin{equation}
\,\,\,\,\,\,\,\,\,\,\,\,\,\,c_1(m) \equiv \frac{b(m)\exp{[b(m)]}}{(2\pi)^m \sqrt{m}}
\label{eq:c1}
\end{equation}

and 

\begin{equation}
\,\,\,\,\,\,\,\,\,\,\,\,\,\,c_2(m) \equiv \left(\frac{b(m)}{2m}\right)^{2m}
\label{eq:c2}
\end{equation}

\vspace{3mm}

we obtain, for $m=1$:

\begin{equation}
\nu_{1}(s) = 2 c_{1}(1) K_{0} (b(1)\,s)
\label{eq:serc_nu1}
\end{equation}

\vspace{3mm}

For $ m= 2  $:

\begin{equation}
2.58697\,s^{-\frac{1}{2}}\,G(\{ \{ \} ,\{ \} \} ,
 \{ \{ 0,\frac{1}{4},\frac{1}{4},\frac{1}{2}\} ,\{ \} \} , 
0.710231\,s^2)
\label{eq:serc_nu2}
\end{equation}

$$
~~~~  \ldots
$$

For $ m= 4  $:

\begin{equation}
 5.26885\,s^{-\frac{3}{4}}\,G(\{ \{ \} ,\{ \} \} , \{ \{ 
    0,\frac{1}{8},\frac{1}{4},\frac{3}{8},\frac{3}{8},\frac{1}{2}, 
 \frac{5}{8},\frac{3}{4}\} ,\{ \} \} ,0.713351\,s^2)
\label{eq:serc_nu4}
\end{equation}

$$
~~~~\ldots
$$
Notice that Eq. (\ref{eq:serc_nu4}) is equivalent to Eq.
(\ref{eq:devauc_nu}) for the \DV\ profile given before.  This may be
seen by applying the identity Eq. (\ref{eq:prop1}) and noting that
$\beta_{S} = \beta_{DV} -1/2 $,
where $\beta_{S}$  and $\beta_{DV}$ denote  the arrays of $\beta$ 
coefficients appearing
respectively in Eq. (\ref{eq:serc_nu4}) and Eq. (\ref{eq:devauc_nu}).

\vspace{3mm}

The expressions above have the general form:

\begin{eqnarray}
\nu_{m}(s) & = &c_{1}(m)\,s^{\frac{1-m}{m}} G(\{ \{ \},\{ \} 
\},\{\{\beta_{S}(m)\},\{ \} \} ,c_{2}(m)\,s^2)  \nonumber   \\
& \equiv  & c_{1}(m)\,s^{\frac{1-m}{m}} 
G^{2m,0}_{0,2m}\left(c_{2}(m)\,s^2 
\big|^{~~~\{~~\}~~,\{~\}}_{\{\beta_{S}(m)\},\{~\}}\right)
\label{eq:serc_nu_gen}
\end{eqnarray}
with $\beta_{S}(m)$ denoting the $2m$-array:

\begin{equation}
\beta_{S}(m) \equiv \left\{~\left(\frac{j -1}{2m}\right)_{1\leq j \leq m} ; 
\left(\frac{j -2}{2m}\right)_{m + 1\leq j \leq 2m} \right\}
\label{eq:beta-ser}
\end{equation}

\vspace{3mm}

\subsection{Other related quantities}

Related quantities  as the mass , the gravitational potential, the (total) 
potential energy and the 
(central) velocity dispersion, can
then  be formally calculated by other integrations, similar to  what
has been done for the \DV\ profile.  From Eq. (\ref{eq:def_mass}) we find for the mass:

\begin{eqnarray}
M(s) & = & 2 \pi  c_{1}(m)\,s^{\frac{2m+1}{m}} 
G\left(\{\{-\frac{1}{2m}\},\{ \} \},\{\{\beta_{S}(m)\},\{ 
-\frac{2m+1}{2m}\}\} ,c_{2}(m)\,s^2 \right)  \nonumber   \\
\nonumber \\
& \equiv & 2 \pi  c_{1}(m)\,s^{\frac{2m+1}{m}} 
G^{2m,1}_{1,2m+1}\left( 
c_{2}(m)\,s^2\bigg|^{\{-\left(\frac{1}{2m}~\right)\},~~~~~\{~~\}}_{\{\beta_{S}(m)\}, 
\{-\left(\frac{2m+1}{2m}\right)\}} \right)
\label{eq:mass_ser}
\end{eqnarray}

\vspace{2mm}

Notice that in the case $m = 1$, because of Eq. (\ref{eq:serc_nu1}), an alternative expression
for the mass may be given by:

\begin{eqnarray}
M(s) & = &  8\,\pi c_1(1) \int _{0}^{s} {\,z^2\,K_0(b \cdot z) d z}  = \nonumber \\
& = & \frac{8\,\pi\, c_1(1)}{18}\, \,s^3\, \Bigg [\, {}_2F_3\left( \{ \frac{3}{2},
\frac{3}{2} \},\{ 1,\frac{5}{2},\frac{5}{2} \} , \frac{b^2 s^2}{4} \right) +  \nonumber \\ 
& - & \frac{3}{2}\, {}_1F_2\left( \{ \frac{3}{2} \},\{ 1,\frac{5}{2} \},\frac{b^2 s^2}{4}
\right) \cdot \log \left( \frac{b^2 s^2}{4} \right) + \frac{9}{2}\,\sum _{j=0}^{\infty }{ 
\frac{2^{1-2 j}\,b^{2 j}\,s^{2 j}\,\Psi(0,1+j)}{(3+2 j)\,{\Gamma^2}(1+j)}}\,\Bigg] 
\label{eq:mass_n=1}
\end{eqnarray}
in which $b$ stands for $b(1)$ and $\Gamma(x)$ and $\Psi(n,x)$  denote 
the gamma function 
and its $(n+1)^{th}$ derivative (i.e, the digamma~-~or {\sl{psi}}~-~function in the case 
$n=0$, and polygamma function in the general case). Both expressions, Eq. (\ref{eq:mass_ser})
and  Eq. (\ref{eq:mass_n=1}) have similar performances in {\sl{Mathematica}}. The total
mass, $M(\infty) = \frac{2\pi }{b} \exp b$, is obtained  to within $10^{-4 }$, for $s = 10$.

\vspace{5mm}

For the gravitational potential, using Eqs. (\ref{eq:def_grav_pot}) and (\ref{eq:mass_ser}), we 
find the following expression:

\begin{eqnarray}
\Psi(s) & = & \pi c_1(m)\,s^{\frac{m+1}{m}} \ast \nonumber \\
& & 
G\left(\{\{-\frac{1}{2m},\frac{m-1}{2m}\},\{\}\},\{\{\beta_{S}(m)\},\{-\frac{2m+1}{2m},-\frac{m+1}{2m}\}\},c_{2}(m)\,s^2\right)  
\nonumber   \\
\nonumber \\
& \equiv & \pi c_1(m)\,s^{\frac{m+1}{m}} 
G^{~2m,2}_{2,2m+2}\left(c_{2}(m)\,s^2 
\bigg|^{\left\{-\left(\frac{1}{2m}~\right), 
\left(\frac{m-1}{2m}\right)\right\},\left\{~~\right\}}_{\left\{\beta_{S}(m)\right\},\left\{-\left(\frac{2m+1}{2m}\right), -\left(\frac{m+1}{2m}\right)\right\}} \right)
\label{eq:psi_ser}
\end{eqnarray}

\vspace{5mm}

Notice that, as before with the equations for
$\nu(s)$ and for the same reasons, in the case $m = 4$, Eqs.
(\ref{eq:mass_ser}) and (\ref{eq:psi_ser}) above will differ from 
those given before in Section 4.1 (Eqs. (\ref{eq:devauc_M}) and
(\ref{eq:pot_dv})).

The gravitational potential energy is defined by:

\begin{equation}
\Omega(s) = \frac{1}{2} \int_{0}^{s}{ \Psi(s) dM(s)} =\frac{1}{2} \int_{0}^{s}{ s^2 \nu(s)\Psi(s) ds}
\label{eq:def_en_pot}
\end{equation} 

Unfortunately there seems to be no formal solution for this integral
in terms of {\sl{Meijer}} functions. However, by making use of a classical
integral of a product of  Meijer functions given in http://functions.wolfram.com/07.09.16.0025
 one finds the following expression for the \textit{total} potential energy:

\begin{eqnarray}
&&\!\!\!\!\!\Omega(\infty)  =\frac{\pi c_1^2(m)}{4\, c_2^{\frac{2 + 3m}{2m}}(m)} \cdot \nonumber \\
\nonumber \\
&& G\left( \left\{\{-\frac{1}{2m}, \frac{1}{2}-\frac{1}{2m},\,  \alpha_E(m) \},\{\}\right\},
\left\{\{\beta_S(m) \},\{-1- \frac{1}{2m},-1/2 -\frac{1}{2m}\}\right\},1\right) \equiv \nonumber \\
\nonumber \\
&& \equiv  \frac{\pi c_1^2(m)}{4\, c_2^{\frac{2 + 3m}{2m}}(m)} \cdot  G^{2m, 2m+2}_{2m+2, 2m+2}\left(1\Big{|}^{\{-\frac{1}{2m}, \frac{1}{2}-\frac{1}{2m},\,  \alpha_E(m) \},\{\}}_
{\{\beta_S(m) \},\{-1- \frac{1}{2m},-1/2 -\frac{1}{2m}\}}\right)
\label{eq:en_pot_tot}
\end{eqnarray}
where, besides the $2m-$array of coefficients $\beta_S(m)$ defined by Eq. (\ref{eq:beta-ser}), 
we have also defined the $2m-$array: $\alpha_E(m) \equiv  -(m+1)/m -\beta_S(m)$. 

The velocity dispersion of a spherical system  in hydrostatic equilibrium is given by:

\begin{equation}
\sigma^2(s) = \frac{1}{\nu(s)} \int^{\infty}_{s}{\frac{M(x) \nu(x)}{x^2} dx}
\label{eq:def_sigma}
\end{equation}

As for the gravitational potential energy, for systems endowed with a \Se\ density
profile, this integral may be expressible in terms of {\sl{Meijer}} functions for the
case $s = 0$. We proceed similarly as to Eq. (\ref{eq:en_pot_tot})  to get:

\begin{eqnarray}
\sigma^2(0) & = &\frac{1}{\nu(0)} \frac{\pi \, c^2_1( m)}{c^{1/2m}_2( m)} \cdot  G\left( \left\{\{-\frac{1}{2m},\, \alpha_{\sigma}(m) \},\{\}\right\},
\left\{\{\beta_S(m) \},\{-1- \frac{1}{2m}\}\right\},1\right) \equiv \nonumber \\
\nonumber \\
& \equiv &\frac{1}{\nu(0)} \frac{\pi \, c^2_1( m)}{c^{1/2m}_2( m)} \cdot   G^{2m, 2m+1}_{2m+1, 2m+1}\left(1\Big{|}^{\{-\frac{1}{2m},\,  \alpha_{\sigma}(m) \},\{\}}_
{\{\beta_S(m) \},\{-1- \frac{1}{2m}\}}\right)
\label{eq:sigma_0}
\end{eqnarray}
with  $\alpha_{\sigma}(m) \equiv  (2m-1)/2m - \beta_S(m)$. 

\section{Conclusions}

   \begin{enumerate}
      \item We obtain analytical solutions for the de-projected \DV\ 
and 
      \Se\ laws using formal integration with {\sl{Mathematica}} as 
well 
      as
      for other related quantities like the mass or the potential, total potential energy and the central
velocity dispersion.

      \item Comparisons with existing numerical estimates show very 
      few differences, however   analytical expressions are 
      always much more 
      very convenient to deal with in many cases.   
   \end{enumerate}

\begin{acknowledgements}
This work was supported by the Brazil-France  CNRS-CNPq cooperation.

We  warmly thank the referee, Dr. Mathews Colless,  for very useful 
comments.
\end{acknowledgements}


\appendix

\section{}

The projected luminosity profile is defined by:

\begin{equation}
L(R) \equiv 2\pi \int_{0}^{R} I(R')R'\,dR' \equiv 2\pi I_0 R_{e}^2 
\int_{0}^{x = R/R_{e}} i(x')x'\,dx'
\label{eq:lumproj}
\end{equation}

\vspace{2mm}

In terms of the dimensionless quantities, $x \equiv R/R_e$ and $l(x) 
\equiv L/I_0 R_{e}^2$, this gives:

\begin{equation}
l(x) = \frac{2m\pi e^{b}}{b^{2m}}\,\gamma(2m,bx^{\frac{1}{m}}) 
\equiv \frac{2m\pi e^{b}}{b^{2m}}\,\left[\Gamma(2m) - 
\Gamma(2m,bx^{\frac{1}{m}})\right]
\end{equation}
%
%
where $\gamma(a,x)$ is the incomplete gamma function and $\Gamma$ its 
complement (cf. Gradshteyn \& Ryzhik, 1980). For integer values of 
$m$ it can also be expressed as:
\vspace{2mm}
\begin{eqnarray}
&&\!\!\!\!\!l(x) = \frac{2m\pi e^{b}}{b^{2m}} \cdot \nonumber \\
&&\;\;\;\left[(2m-1)! - \exp{(-bx^{1/m})}\left((2m-1)!
+ \sum_{j=1}^{2m-1} \frac{(2m-1)!}{(2m-j)!} b^{2m-j}x^{(2m-j)/m}
\right)\right]
\end{eqnarray}

\vspace{3mm}

From this one may find $b(m)$ as the solutions of the equation 
$L(R_e) \equiv L_{tot}/2$, where $L_{tot}$ is the total luminosity 
integrated to infinity. We have (see also Ciotti  1991):

\begin{equation}
\gamma(2m,b) \equiv \Gamma(2m,b) = \Gamma(2m)/2
\label{eq:bm}
\end{equation}

This is solved instantaneously using the following  {\sl Mathematica} 
commands: 


\begin{verbatim}    
   mm = {1, 2, 3, 4, 5, 6, 7, 8, 9, 10, 11, 12, 13, 14, 15}    
   b[m_] := FindRoot[Gamma[2m, b] == Gamma[2m]/2, {b, 2m - 1/3},
         WorkingPrecision -> 60, AccuracyGoal -> 30]
	 blist = Map[b, mm]; bb = N[b /. blist, 17]    
\end{verbatim}

The derived values of $b(m)$ are given in Table \ref{bvalues},
where we compare the {\sl{Mathematica}} results with those from
the asymptotic expansions by Ciotti \& Bertin (1999).
For $m=4$, we find of course $b(4) = 7.66925$. 

\begin{table}[htb]
\caption[]{The values of $b(m)$}
\tabcolsep=0.6\tabcolsep
\begin{tabular}{r c c l }
\hline
$m$ &      $b(m)$       & $b_{CB99}$      & 
~~~~~~~~$b(m)-b_{CB99}$    \\
\hline
1   &  1.67834699001666 &1.67838865492157  & 
-4.16649049 $10^{-5} $  \\

2   &  3.67206074885089 &3.67206544591768  & 
-4.6970667 $10^{-6} $  \\

3   &  5.67016118871207 &5.67016250849902  & 
-1.3197869   $10^{-6} $  \\

4   &  7.6692494425008  &7.66924998466950  & 
-5.421687   $10^{-7} $  
\\
5   &  9.66871461471413 &9.66871488808778  & 
-2.733736  $10^{-7} $   \\

6   & 11.6683631530448  &11.6683633097115  & 
-1.566667 $10^{-7} $   \\

7   & 13.6681145993449  &13.6681146973462  & 
-9.80013  $10^{-8} $    \\

8   & 15.6679295443172  &15.6679296096535  & 
-6.53363  $10^{-8} $    \\

9   & 17.6677864177885  &17.6677864635090  & 
-4.57206 $10^{-8} $    \\

10  & 19.6676724233057  &19.6676724565414  & 
-3.32357  $10^{-8} $    \\

11  & 21.6675794898319  &21.6675795147457  & 
-2.49138 $10^{-8} $    \\

12  & 23.6675022752263  &23.6675022943807  & 
-1.91544  $10^{-8} $    \\

13  & 25.6674371029624  &25.6674371180047  & 
-1.50423  $10^{-8} $    \\

14  & 27.6673813599995  &27.6673813720274  & 
-1.20280  $10^{-8} $    \\

15  & 29.6673331382212  &29.6673331479896  &
-9.7684 $10^{-9} $    \\
\hline
\label{bvalues}
\end{tabular}
\end{table}

\vspace{3mm}

The luminosity so writes:
  
\vspace{3mm}

$\bullet$ for $m=2$, with $ b=b(2)$:

\begin{equation}
\frac{4\,\pi}{b^4}\,e^b \,[\,6 - e^{- b\,{\sqrt{x}}}\,
(6 + 6\,b\,{\sqrt{x}} + 3\,b^2\,x + b^3\,x^{\frac{3}{2}}\,)]
\end{equation}

\vspace{3mm}

$\bullet$ for $m=4$, with $ b=b(4)$:

\begin{eqnarray}
\!\!\!\!\!&\frac{8\,\pi}{b^8}\,e^b \,[\,5040 & - \nonumber \\
& &e^{- b\,x^{\frac{1}{4}}}\,(5040 + 
5040\,b\,x^{\frac{1}{4}} + 2520\,b^2\,{\sqrt{x}} + 
840\,b^3\,x^{\frac{3}{4}} + \nonumber \\
& &210\,b^4\,x + 42\,b^5\,x^{\frac{5}{4}} + 
7\,b^6\,x^{\frac{3}{2}} + b^7\,x^{\frac{7}{4}})]
\end{eqnarray}

\vspace{3mm}

~~~~~ and so on $\ldots$


\begin{thebibliography}{}

\bibitem[CiottiA, 1991]{ciottiA} Ciotti, L., 1991, Astron. \& 
Astrophys. 249, 99
    
\bibitem[Ciotti \& Bertin, 1999]{ciottiB} Ciotti, L., Bertin, G., 
1999, Astron. \& Astrophys. 353, 447
    
\bibitem[de Vaucouleurs, 1948]{devauc48} de Vaucouleurs, G., 1948,  
Ann.d'Astroph. 11, 247
       
\bibitem[Fuchs \& Materne, 1982]{mat82} Fuchs, B., Materne, J., Astron. 
\& Astrophys. , 113, 85.
 
\bibitem[Gradshteyn \& Ryzhik, 1980]{grad80} Gradshteyn, I.S. 
and Ryzhik, I.M 1980,{\textit{Table of Integrals, Series and 
Products}},
Academic Press.

\bibitem[Graham \& Colless, 1997]{graham97}  Graham, A., Colless, 
M.,  
1997, Mon. Not. R. Astron. Soc. 287, 221
  
\bibitem[Marquez, 2001]{marquez} Marquez, I., Lima Neto, G.B., 
Capelato, H.V., Durret, F., Lanzoni, B., Gerbal, D., 2001, Astron.\& 
Astrophys. in press (Astro-ph 0009474).

\bibitem[Mellier \& Mathez, 1987]{mm87} Mellier, Y., Mathez, G., 
1987,  Astron.\& Astrophys., 175,1.

\bibitem[Poveda et al, 1960]{Poveda} Poveda, A.,  Iturriaga, R., and 
Orozco, I., 1960, Bol. Obs. Tonantzinla Nƒ 20, P. 3
  
\bibitem[S\'ersic, 1968]{Sersic} S\'ersic, J.-L.,  1968, 
{\textit{Atlas de 
Galaxias Australes}}, Observatorio Astronomico de Cordoba.
 
\bibitem[Young, 1976]{Young} Young, P. J., 1976,  AJ, 81, 807

\bibitem[Wolfram, 1991]{mathbook} Wolfram, S., 1991, {\textit{The
 Mathematica Handbook}}, Addison-Wesley Publishing Company, Inc.

\end{thebibliography}
\end{document}